\input harvmac
\input amssym
\def\p{\partial}

\def\R{R}
\noblackbox

\lref\sct{A useful book with references to the extensive original literature is J.F. Futterman,
F. Handler and R. Matzner, ``Scattering from Black Holes'', Cambridge University Press
(1988).}
\lref\masu{J.~ M.~ Maldacena and L. ~Susskind, ``D-branes and Fat Black Holes,'' Nucl. Phys. {\bf B 475},
679 (1996).}
\lref\ReggeZD{
 T.~Regge and C.~Teitelboim,
 ``Role Of Surface Integrals In The Hamiltonian Formulation Of General
 Relativity,''
 Annals Phys.\  {\bf 88}, 286 (1974).
}

\lref\BrownNW{
 J.~D.~Brown and M.~Henneaux,
 ``Central Charges in the Canonical Realization of Asymptotic Symmetries: An
 Example from Three-Dimensional Gravity,''
 Commun.\ Math.\ Phys.\  {\bf 104}, 207 (1986).

}

\lref\BarnichJY{
 G.~Barnich and F.~Brandt,
 ``Covariant theory of asymptotic symmetries, conservation laws and  central
 charges,''
 Nucl.\ Phys.\  B {\bf 633}, 3 (2002)
 [arXiv:hep-th/0111246].
}

\lref\BarnichBF{
 G.~Barnich and G.~Compere,
 ``Surface charge algebra in gauge theories and thermodynamic integrability,''
 J.\ Math.\ Phys.\  {\bf 49}, 042901 (2008)
 [arXiv:0708.2378 [gr-qc]].
}

\lref\PressZZ{
  W.~H.~Press and S.~A.~Teukolsky,
  ``Perturbations of a Rotating Black Hole. II. Dynamical Stability of the Kerr
  Metric,''
  Astrophys.\ J.\  {\bf 185}, 649 (1973).
}

\lref\TeukolskyHA{
  S.~A.~Teukolsky,
  ``Perturbations of a rotating black hole. I. Fundamental equations for
  gravitational electromagnetic and neutrino field perturbations,''
  Astrophys.\ J.\  {\bf 185}, 635 (1973).
}

\lref\TeukolskyYV{
  S.~A.~Teukolsky and W.~H.~Press,
  ``Perturbations Of A Rotating Black Hole. III - Interaction Of The Hole With
  Gravitational And Electromagnet Ic Radiation,''
  Astrophys.\ J.\  {\bf 193}, 443 (1974).
}

\lref\GuicaMU{
  M.~Guica, T.~Hartman, W.~Song and A.~Strominger,
  ``The Kerr/CFT Correspondence,''
  arXiv:0809.4266 [hep-th].
}

\lref\BardeenPX{
  J.~M.~Bardeen and G.~T.~Horowitz,
  ``The extreme Kerr throat geometry: A vacuum analog of AdS(2) x S(2),''
  Phys.\ Rev.\  D {\bf 60}, 104030 (1999)
  [arXiv:hep-th/9905099].
}

\lref\EmparanEN{
  R.~Emparan and A.~Maccarrone,
  ``Statistical Description of Rotating Kaluza-Klein Black Holes,''
  Phys.\ Rev.\  D {\bf 75}, 084006 (2007)
  [arXiv:hep-th/0701150].
}

\lref\DiasNJ{
  O.~J.~C.~Dias, R.~Emparan and A.~Maccarrone,
  ``Microscopic Theory of Black Hole Superradiance,''
  Phys.\ Rev.\  D {\bf 77}, 064018 (2008)
  [arXiv:0712.0791 [hep-th]].
}

\lref\AmselEV{
  A.~J.~Amsel, G.~T.~Horowitz, D.~Marolf and M.~M.~Roberts,
  ``No Dynamics in the Extremal Kerr Throat,''
  arXiv:0906.2376 [hep-th].
}
\lref\ascv{A. Strominger and C. Vafa, ``Microscopic Origin of the Bekenstein-Hawking Entropy,''
Phys. Lett. {\bf B 379}, 99 (1996) [hep-th/9601029].}
\lref\nhbh{A. Strominger, ``Black hole entropy from near-horizon microstates,'' JHEP {\bf 9802}, 009
(1998) [arXiv:hep-th/9712251].}
\lref\DiasEX{
  O.~J.~C.~Dias, H.~S.~Reall and J.~E.~Santos,
  ``Kerr-CFT and gravitational perturbations,''
  arXiv:0906.2380 [hep-th].
}

\lref\MaldacenaBW{
  J.~M.~Maldacena and A.~Strominger,
  ``AdS(3) black holes and a stringy exclusion principle,''
  JHEP {\bf 9812}, 005 (1998)
  [arXiv:hep-th/9804085].
}

\lref\MaldacenaIH{
  J.~M.~Maldacena and A.~Strominger,
  ``Universal low-energy dynamics for rotating black holes,''
  Phys.\ Rev.\  D {\bf 56}, 4975 (1997)
  [arXiv:hep-th/9702015].
}

\lref\CveticUW{
  M.~Cvetic and F.~Larsen,
  ``General rotating black holes in string theory: Greybody factors and  event
  horizons,''
  Phys.\ Rev.\  D {\bf 56}, 4994 (1997)
  [arXiv:hep-th/9705192].
}

\lref\CveticXV{
  M.~Cvetic and F.~Larsen,
  ``Greybody factors for rotating black holes in four dimensions,''
  Nucl.\ Phys.\  B {\bf 506}, 107 (1997)
  [arXiv:hep-th/9706071].
}

\lref\CveticVP{
  M.~Cvetic and F.~Larsen,
  ``Black hole horizons and the thermodynamics of strings,''
  Nucl.\ Phys.\ Proc.\ Suppl.\  {\bf 62}, 443 (1998)
  [Nucl.\ Phys.\ Proc.\ Suppl.\  {\bf 68}, 55 (1998)]
  [arXiv:hep-th/9708090].
}
\lref\CveticAP{
  M.~Cvetic and F.~Larsen,
  ``Greybody factors for black holes in four dimensions: Particles with
  spin,''
  Phys.\ Rev.\  D {\bf 57}, 6297 (1998)
  [arXiv:hep-th/9712118].
}

\lref\MaldacenaUZ{
  J.~M.~Maldacena, J.~Michelson and A.~Strominger,
  ``Anti-de Sitter fragmentation,''
  JHEP {\bf 9902}, 011 (1999)
  [arXiv:hep-th/9812073].
}
\lref\BredbergPV{
  I.~Bredberg, T.~Hartman, W.~Song and A.~Strominger,
  ``Black Hole Superradiance From Kerr/CFT,''
  arXiv:0907.3477 [hep-th].
}

\lref\RasmussenIX{
  J.~Rasmussen,
  ``Isometry-preserving boundary conditions in the Kerr/CFT correspondence,''
  arXiv:0908.0184 [hep-th].
}

\lref\AmselPU{
  A.~J.~Amsel, D.~Marolf and M.~M.~Roberts,
  ``On the Stress Tensor of Kerr/CFT,''
  arXiv:0907.5023 [hep-th].
}

\lref\CastroJF{
  A.~Castro and F.~Larsen,
  ``Near Extremal Kerr Entropy from AdS$_2$ Quantum Gravity,''
  JHEP {\bf 0912}, 037 (2009)
  [arXiv:0908.1121 [hep-th]].
}

\lref\MatsuoSJ{
  Y.~Matsuo, T.~Tsukioka and C.~M.~Yoo,
  ``Another Realization of Kerr/CFT Correspondence,''
  arXiv:0907.0303 [hep-th].
}

\lref\MatsuoPG{
  Y.~Matsuo, T.~Tsukioka and C.~M.~Yoo,
  ``Yet Another Realization of Kerr/CFT Correspondence,''
  arXiv:0907.4272 [hep-th].
}

\lref\carter{B. Carter, Phys. Rev. {\bf 174} 1559; Comm. Math. Phys. {\bf 10} 280.}
\lref\HartmanNZ{
  T.~Hartman, W.~Song and A.~Strominger,
  ``Holographic Derivation of Kerr-Newman Scattering Amplitudes for General
  Charge and Spin,''
  arXiv:0908.3909 [hep-th].
}

\lref\CveticJN{
  M.~Cvetic and F.~Larsen,
  ``Greybody Factors and Charges in Kerr/CFT,''
  JHEP {\bf 0909}, 088 (2009)
  [arXiv:0908.1136 [hep-th]].
}

\lref\BredbergPV{
  I.~Bredberg, T.~Hartman, W.~Song and A.~Strominger,
  ``Black Hole Superradiance From Kerr/CFT,''
  arXiv:0907.3477 [hep-th].
}

\lref\gen{
  H.~Lu, J.~Mei and C.~N.~Pope,
  ``Kerr/CFT Correspondence in Diverse Dimensions,''
  JHEP {\bf 0904}, 054 (2009)
  [arXiv:0811.2225 [hep-th]].
 \vskip 0.08mm \noindent
  T.~Azeyanagi, N.~Ogawa and S.~Terashima,
  ``Holographic Duals of Kaluza-Klein Black Holes,''
  JHEP {\bf 0904}, 061 (2009)
  [arXiv:0811.4177 [hep-th]].
\vskip 0.08mm \noindent
  T.~Hartman, K.~Murata, T.~Nishioka and A.~Strominger,
  JHEP {\bf 0904}, 019 (2009)
  [arXiv:0811.4393 [hep-th]].
  \vskip 0.08mm \noindent
  D.~D.~K.~Chow, M.~Cvetic, H.~Lu and C.~N.~Pope,
  ``Extremal Black Hole/CFT Correspondence in (Gauged) Supergravities,''
  arXiv:0812.2918 [hep-th].
\vskip 0.08mm \noindent
  H.~Isono, T.~S.~Tai and W.~Y.~Wen,
  ``Kerr/CFT correspondence and five-dimensional BMPV black holes,''
  arXiv:0812.4440 [hep-th].
\vskip 0.08mm \noindent
  T.~Azeyanagi, N.~Ogawa and S.~Terashima,
  ``The Kerr/CFT Correspondence and String Theory,''
  arXiv:0812.4883 [hep-th].
\vskip 0.08mm \noindent
  J.~J.~Peng and S.~Q.~Wu,
  ``Extremal Kerr black hole/CFT correspondence in the five dimensional G\'odel
  universe,''
  Phys.\ Lett.\  B {\bf 673}, 216 (2009)
  [arXiv:0901.0311 [hep-th]].
\vskip 0.08mm \noindent
  C.~M.~Chen and J.~E.~Wang,
  ``Holographic Duals of Black Holes in Five-dimensional Minimal
  Supergravity,''
  arXiv:0901.0538 [hep-th].
\vskip 0.08mm \noindent
  F.~Loran and H.~Soltanpanahi,
  ``5D Extremal Rotating Black Holes and CFT duals,''
  arXiv:0901.1595 [hep-th].
  \vskip 0.08mm \noindent
  A.~M.~Ghezelbash,
  ``Kerr/CFT Correspondence in Low Energy Limit of Heterotic String Theory,''
  arXiv:0901.1670 [hep-th].
\vskip 0.08mm \noindent
  H.~Lu, J.~w.~Mei, C.~N.~Pope and J.~F.~Vazquez-Poritz,
  ``Extremal Static AdS Black Hole/CFT Correspondence in Gauged
  Supergravities,''
  Phys.\ Lett.\  B {\bf 673}, 77 (2009)
  [arXiv:0901.1677 [hep-th]].
  \vskip 0.08mm \noindent
  G.~Compere, K.~Murata and T.~Nishioka,
  ``Central Charges in Extreme Black Hole/CFT Correspondence,''
  JHEP {\bf 0905}, 077 (2009)
  [arXiv:0902.1001 [hep-th]].
\vskip 0.08mm \noindent
  K.~Hotta,
  ``Holographic RG flow dual to attractor flow in extremal black holes,''
  arXiv:0902.3529 [hep-th].
\vskip 0.08mm \noindent
  D.~Astefanesei and Y.~K.~Srivastava,
  ``CFT Duals for Attractor Horizons,''
  arXiv:0902.4033 [hep-th].
\vskip 0.08mm \noindent
  A.~M.~Ghezelbash,
  ``Kerr-Bolt Spacetimes and Kerr/CFT Correspondence,''
  arXiv:0902.4662 [hep-th].
\vskip 0.08mm \noindent
  M.~R.~Garousi and A.~Ghodsi,
  ``The RN/CFT Correspondence,''
  arXiv:0902.4387 [hep-th].
\vskip 0.08mm \noindent
  T.~Azeyanagi, G.~Compere, N.~Ogawa, Y.~Tachikawa and S.~Terashima,
  ``Higher-Derivative Corrections to the Asymptotic Virasoro Symmetry of 4d
  Extremal Black Holes,''
  arXiv:0903.4176 [hep-th].
\vskip 0.08mm \noindent
  X.~N.~Wu and Y.~Tian,
  ``Extremal Isolated Horizon/CFT Correspondence,''
  arXiv:0904.1554 [hep-th].
\vskip 0.08mm \noindent
  J.~Rasmussen,
  ``Isometry-preserving boundary conditions in the Kerr/CFT correspondence,''
  arXiv:0908.0184 [hep-th].

}

\Title{\vbox{\baselineskip12pt
}}
{\vbox{\centerline {Hidden Conformal Symmetry  }
\centerline {of the Kerr Black Hole} }}


\centerline{
Alejandra Castro${}^{\diamond}$, Alexander Maloney${}^{\diamond}$ and Andrew Strominger${}^{\dagger}$}

\bigskip
\centerline{${}^\diamond$\it Physics Department, McGill University, Montreal,
CA}
\smallskip
\centerline{${}^\dagger$\it Center for the Fundamental Laws of Nature, Harvard University, Cambridge, MA, USA}
\smallskip

\vskip .3in
\centerline{\bf Abstract}
\smallskip

    Extreme and very-near-extreme spin $J$ Kerr black holes have been  conjectured  to be
 holographically dual to two-dimensional (2D) conformal field theories (CFTs) with left and right central charges
 $c_L=c_R=12J$. In this paper it is observed that the 2D conformal symmetry of the scalar wave equation at low frequencies persists for generic  non-extreme values of the mass $M\neq \sqrt{J}$. Interestingly, this conformal  symmetry  is $not$ derived from a conformal symmetry of the spacetime geometry except in the extreme limit.  The $2\pi$ periodic identification of the azimuthal angle $\phi$ is shown to correspond to a spontaneous breaking of the conformal symmetry by  left and right temperatures $T_L=M^2/2\pi J$ and $T_R=\sqrt{M^4-J^2}/2\pi J$. The well-known low-frequency scalar-Kerr scattering amplitudes coincide with correlators  of a  2D CFT at these temperatures.  Moreover the CFT microstate degeneracy inferred from the Cardy formula agrees exactly with the Bekenstein-Hawking area law for all $M$ and $J$. These observations provide evidence for  the conjecture that the  Kerr black hole is dual to a $c_L=c_R=12J$ 2D CFT at temperatures $(T_L,T_R)$ for every value of $M$ and $J$.

\Date{}
\listtoc
\writetoc

\newsec{Introduction}

An extreme  Kerr black hole with mass $M$ and angular momentum $J=M^2$ has a near-horizon scaling region, known as the NHEK (Near-Horizon Extreme Kerr) geometry, which has an enhanced $SL(2,R)\times U(1)$ isometry group \BardeenPX. Recently it has been shown \refs{\GuicaMU, \MatsuoSJ,\CastroJF}   from an analysis of the NHEK boundary conditions that the canonically conserved charges associated with the non-trivial diffeomorphisms of the NHEK region form two copies of the two-dimensional Virasoro algebra. The central charges were computed to be $c_L=c_R=12J$.
This motivated the conjecture \GuicaMU\ that the extreme Kerr black hole is dual to a two-dimensional CFT.  The conjecture was supported by the facts that, at and very near extremality, the Cardy CFT microstate degeneracy precisely matches the Bekenstein-Hawking entropy and the
finite temperature CFT correlators precisely match the Kerr scattering amplitudes.  Other tests of the Kerr/CFT conjecture and its generalizations, all successful, appear in \refs{\gen,\BredbergPV, \CveticJN, \HartmanNZ}.

If the conjecture is correct finite excitations of the CFT are expected to correspond to generic
non-extremal Kerr black holes.  However all attempts so far to understand Kerr black holes in this manner a finite distance from the extreme limit have run into obstacles. The problem is that away from the extreme limit the NHEK geometry disappears and the near-horizon geometry is just Rindler space. We know of no clear way to associate a conformal field theory to Rindler space. Put another way, the back reaction of a finite energy excitation on the geometry appears to destroy the conformal symmetry. This is closely related to the AdS$_2$ fragmentation problem discussed in \MaldacenaUZ.

The key observation of the present paper, which enables us to circumvent this obstacle, is that a near horizon geometry (such as NHEK or AdS$_3$) with a conformal symmetry group is {\it not} a {\it necessary} condition for the interactions to exhibit conformal invariance. For scattering amplitudes,  a {\it sufficient} condition is that the solution space of the wave equation for the propagating field has a conformal symmetry.
Such a symmetry is guaranteed if the space on which the field propagates has the symmetry.    However we will see that it can and does happen that the solution space
has the requisite conformal symmetry even when the space on which the field propagates does not.

While we will see this conformal symmetry emerge  in detail in the text, it is possible to understand heuristically why this occurs.  At low frequencies $\omega \ll {1 \over M}$ the wave equation can be solved
with a matching procedure which divides the geometry into a near region $r\ll {1 \over \omega}$
and a far region $ r \gg  M$ which have a large overlap.  The solution of the full wave equation is obtained by matching the inner part of the far region solution with the outer part of the near region solution along a matching surface $r_M(t,\phi)$. In order for the matching procedure to be consistent, the final result cannot depend on the arbitrary choice of the matching surface $r_M$.  This requires that the amplitudes in each region have a symmetry under arbitrary local changes of $r_M$. Changing $r_M$ changes the redshift factor at
the matching surface, and so is a local change in scale. It is thus perhaps not surprising that this system has a local 2D conformal symmetry.  For the case of extreme Kerr, or for the BPS black holes studied in string theory, the near region turns out -- for special reasons --  to be equivalent to the near horizon region and the conformal symmetry of the wave equation is lifted from the conformal symmetry of the geometry. In the generic case this is not so. The near region goes out to values of $r\ll {1 \over \omega}$ with $r\gg M$
and so essentially includes the entire asymptotically flat spacetime.

The  conformal symmetry we find acts locally on the solution space, but is globally obstructed by periodic identification of the azimuthal angle $\phi$. We argue that this spontaneous breaking of the conformal group is precisely of the form produced by finite left and right temperatures $T_L=M^2/2\pi J$ and $T_R=\sqrt{M^4-J^2}/2\pi J$ in a 2D CFT.  This suggestion is corroborated by the demonstration that the known \sct\ near-region scattering amplitudes computed in the 70s are indeed of the form required by conformal invariance for a finite temperature 2D CFT.
Moreover, using the temperatures $(T_L,T_R)$ and the values of the central charge $c_L=c_R=12J$ previously computed at extremality one can apply the Cardy formula to count the number of states.  This precisely reproduces Bekenstein-Hawking Area law for the black hole entropy
\eqn\yyl{S_{micro}={\pi^2 \over 3}(c_LT_L+c_RT_R)=2\pi (M^2+\sqrt{M^4-J^2})={{\rm Area} \over 4}~.}
These results all support the Kerr/CFT conjecture for general $J$ and $M$.\foot{As mentioned in \GuicaMU\ for extreme Kerr, there may be an underlying "long string" interpretation \masu\ involving the $J$-fold cover of the CFT circle.
The long string has $c_R=c_L=12$ and the temperatures and charges are rescaled by a factor of $J$.
We will not reiterate here the issues surrounding the long string picture but wish to note that it also has appealing features  for the case of general $M$ and $J$ considered here.}

We wish to warn the reader that we have $not$, in this paper, provided a systematic derivation or even argument from some set of assumptions that a generic Kerr black hole
is dual to a 2D CFT. In past examples such derivations have proceeded  from an analysis of the asymptotic symmetry group of the geometry \refs{\BrownNW,\nhbh,\GuicaMU, \MatsuoSJ,\CastroJF}, or from a scaling limit of string theory \ascv. Since the conformal symmetries here are not symmetries of the spacetime geometry, and we are not embedding in string theory, these approaches can not work.\foot{Clearly a new approach is needed. Perhaps there is a generalization of the notion of an asymptotic symmetry group of a dynamical system which does not insist that the symmetries
are purely geometric and allows for the more general realization of conformal symmetry discussed here.} In the absence of a systematic approach we have patched together, and provided evidence for, a picture with what strikes us as a remarkable cohesiveness.  However, holes in the picture remain and we hope to have inspired the reader to fill them in!

This paper is organized as follows.  In section 2 we review the massless scalar wave equation in the Kerr background. In section 3 we describe the near region where the behaviour of this wave equations simplifies.  In section 4 we locally construct six vector fields  with an  $SL(2,\R) \times SL(2,\R)$ Lie bracket algebra, show that their Casimir is precisely the near-region scalar wave equation, and identify them as generators of a conformal symmetry spontaneously broken down to $U(1)\times U(1)$ by the $2\pi$ identification of the azimuthal angle.  A dual CFT interpretation is proposed in section 5, which allows us to compute the left and right CFT temperatures and hence the microscopic entropy using a Cardy formula.  In section 6 we provide further evidence for the proposed generalized Kerr/CFT correspondence by showing that the scattering amplitudes in the near region agree with those of a finite temperature 2D CFT.


\newsec{Massless Scalar Wave Equation}

In this section we describe the classical wave equation for a massless scalar on the geometry
of a Kerr black hole with generic  mass $M$ and angular momentum $J=Ma$.
We use the familiar Boyer-Lindquist coordinates
\eqn\aa{ds^2={\rho^2\over {\Delta}}dr^2- {{\Delta}\over \rho^2}\left(dt -{a}\sin^2\theta d\phi\right)^2+ \rho^2d\theta^2+{ \sin^2\theta \over \rho^2} \left((r^2+{a}^2)d\phi-{{a}}dt\right)^2~,}
where  ${\Delta}$ and $\rho^2$ are given by
\eqn\ac{{\Delta}=r^2 +{a}^2-2M r~,\quad \rho^2=r^2+{a}^2\cos^2\theta~.}
The inner and outer horizons are located at
\eqn\ad{{r}_\pm = M \pm \sqrt{M^2-{a}^2}~.}

The Klein-Gordon equation for a massless scalar  is
\eqn\ba{{1\over \sqrt{-g} } \partial_\mu\left(\sqrt{-g}g^{\mu\nu}\partial_\nu\Phi\right)=0~.}
Expanding in eigenmodes
\eqn\bab{\Phi(t,r,\theta, \phi) = e^{-i\omega t+im\phi}\Phi(r,\theta)~,}
and using \aa\ equation \ba\ becomes
\eqn\bb{\eqalign{&\partial_r\left(\Delta \partial_r \Phi\right)+{\left({2Mr_+\omega}-{a}m\right)^2\over (r-r_+)(r_+-r_-)}\Phi - {\left({2Mr_-\omega}-{a}m\right)^2\over (r-r_-)(r_+-r_-)}\Phi\cr &+\left(r^2+a^2\cos^2\theta+2M(r+2M)\right)\omega^2\Phi+\nabla_{S^2}\Phi=0~.}}
Famously \refs{\carter}, this equation (as well as its higher spin and fermionic cousins \refs{\TeukolskyHA, \PressZZ, \TeukolskyYV}) can be separated.
Writing
\eqn\ep{\Phi(r,\theta)= R(r)S(\theta)~,}
we have
\eqn\bba{ \left[ {1\over \sin\theta}\partial_\theta
\left(\sin\theta\partial_\theta\right)-{m^2\over\sin^2\theta}+\omega^2 a^2\cos^2\theta \right] S(\theta)=-K_\ell  S(\theta)~,}
and
\eqn\bbz{\eqalign{\left[
\partial_r \Delta \partial_r
+{({2Mr_+\omega}-{a}m )^2\over (r-r_+)(r_+-r_-)}   - {({2Mr_-\omega}-{a}m )^2\over (r-r_-)(r_+-r_-)}
 + (r^2+2M(r+2M) )\omega^2 \right] R(r)=K_\ell  R(r)~ .}}
Both equations are solved by Heun functions and the separation constants $K_\ell$ are the eigenvalues on a sphere. The Heun functions are not among the usual special functions and the
$K_\ell$ are known only numerically.


\newsec{The Near Region }

We start by asking whether it is possible to find a range of parameters where the order $\omega^2$ terms in the second line of \bb\ can be neglected; in this case, as we will see below, the wave equation simplifies considerably.  We see from \bb\ that this occurs when the wavelength of the scalar excitation is large compared to the radius of curvature
\eqn\bda{ \omega M\ll 1~.}
In this case the geometry can be divided into two regions
\eqn\wui{\eqalign{& r\ll {1 \over \omega}~~~~~``{\rm NEAR}"\cr
& r\gg M~~~~~``{\rm FAR}"\cr}}
which have significant overlap in the matching region
\eqn\tti{ M\ll r  \ll {1 \over \omega}~~~~~~~``{\rm MATCHING}"~.}
The wave equations in the near and far regions can be solved in terms of familiar special functions, and a full solution is obtained by matching near and far solutions together along a surface in the matching region.

We note that the near region defined above is not the same as the oft-discussed ``near-horizon'' region of the geometry defined by $r-r_+\ll M$.  Indeed, for sufficiently small
$\omega$, the value of $r$ in the near region defined by \wui\ can be arbitrarily large.  For a generic non-extreme Kerr the near-horizon geometry is just Rindler space, while the structure of the near region is   more complicated.

 We view the far region as an asymptotic region where the scattering experiments are set up.  The black hole is thought of as encompassing the whole ``near" region.
 Waves are sent from the far region into the matching region, which is the interface for interactions with the black hole.   We will see that the behaviour of these incident waves in the near region has conformal symmetry.  This conformal invariance results form the freedom to locally choose the radius of the matching surface within the matching region.

In the near region, the angular equation \bda\ reduces to the standard Laplacian on the 2-sphere
\eqn\bea{ \left[ {1\over \sin\theta}\partial_\theta
\left(\sin\theta\partial_\theta\right)-{m^2\over\sin^2\theta} \right]S(\theta)=-K_{\ell}S(\theta)~,\quad \ell =-m,\cdots, m~.}
The solutions $e^{im\phi}S(\theta)$ are spherical harmonics, and the separation constants are
\eqn\fol{K_\ell=\ell(\ell+1)~.}
The radial wave equation in the limit \bda\ becomes\foot{When $m\neq0$ in certain regions of $r$ and/or in the black hole parameter space $(M,J)$ , it is possible to drop in addition the $\omega$ terms in the numerators of the poles in this expression. See subsection 6.2 for further discussion.}
\eqn\bfxx{\eqalign{\left[
\partial_r \Delta \partial_r
+{({2Mr_+\omega}-{a}m )^2\over (r-r_+)(r_+-r_-)}   - {({2Mr_-\omega}-{a}m )^2\over (r-r_-)(r_+-r_-)}
 \right] R(r)=\ell(\ell+1)  R(r)~ .}}
The above equation is solved by hypergeometric functions.  As hypergeometric functions transform in representations of $SL(2,R)$, this suggests the existence of a hidden conformal symmetry. This is the subject of the next section.


\newsec{$SL(2,R)_L\times SL(2,R)_R$}

In this section we will describe the $SL(2,R)_L\times SL(2,R)_R$ symmetry of the
near-region scalar field equation.  For this purpose it is convenient to adapt ``conformal'' coordinates
$(w^\pm,y)$ defined in terms of $(t,r, \phi)$ by

\eqn\cc{\eqalign{w^+ &=\sqrt{r-r_+\over r-r_-}\, e^{2\pi T_R \phi}\cr w^- &=\sqrt{r-r_+\over r-r_-}\,e^{2\pi T_L \phi-{t \over 2M}}
\cr
y &=\sqrt{r_+-r_-\over r-r_-}\,e^{\pi (T_L+T_R) \phi-{t \over 4M}}}}
where
\eqn\llj{T_R\equiv {r_+-r_- \over 4\pi a}~, ~~~~~~~~~~~~T_L\equiv {r_++r_- \over 4\pi a}~.}

Next we define $locally$ the vector fields
\eqn\bg{\eqalign{H_1&=i\p_+~,\cr   H_0&=i(w^+\p_++{1 \over 2}y\p_y)~,\cr
H_{-1}&=i(w^{+2}\p_++w^+y\p_y-y^2\p_- )~,    }}
and
\eqn\bsg{\eqalign{\bar H_1&=i\p_-~,\cr   \bar H_0&=i(w^-\p_-+{1 \over 2}y\p_y)~,\cr
\bar H_{-1}&=i(w^{-2}\p_-+w^-y\p_y-y^2\p_+ )~.       }}
These obey the $SL(2,R)$ Lie bracket algebra,
\eqn\bga{[H_0,H_{\pm 1}]=\mp i H_{\pm 1}~, \quad [H_{-1},H_{1}]=-2iH_0~,}
and similarly for $(\bar H_0, \bar H_{\pm1})$. The $SL(2,R)$ quadratic Casimir is
\eqn\bgb{\eqalign{{\cal H}^2&=\bar { \cal H}^2=-H_0^2+{1\over 2}(H_1H_{-1}+H_{-1}H_1)\cr&={1 \over 4}(y^2\p_y^2-y\p_y )+y^2\p_+\p_-~.}}

In terms of the $ (t, r, \phi)$ coordinates, the vector fields are
\eqn\bzg{\eqalign{H_1&=ie^{-2\pi T_R \phi}\left(\Delta^{1/2}\partial_r+{1\over 2\pi T_R}{r-M\over \Delta^{1/2} }\partial_\phi+{2T_L\over T_R}{Mr-a^2\over \Delta^{1/2}}\partial_t\right)~,\cr   H_0&={i \over 2\pi T_R}\p_\phi +2iM{T_L \over T_R}\p_t~,\cr
H_{-1}&=ie^{2\pi T_R \phi}\left(-\Delta^{1/2}\partial_r+{1\over 2\pi T_R}{r-M\over \Delta^{1/2} }\partial_\phi+{2T_L\over T_R}{Mr-a^2\over \Delta^{1/2}}\partial_t\right)~, \cr      }}
and
\eqn\bsga{\eqalign{\bar H_1&=ie^{-2\pi T_L \phi+{t \over 2M}}\left(\Delta^{1/2}\partial_r-{a\over \Delta^{1/2} }\partial_\phi-2M{r\over \Delta^{1/2}}\partial_t\right)~,\cr   \bar H_0&=-2iM\p_t~,\cr
\bar H_{-1}&=ie^{2\pi T_L \phi-{t \over 2M}}\left(-\Delta^{1/2}\partial_r-{a\over \Delta^{1/2} }\partial_\phi-2M{r\over \Delta^{1/2}}\partial_t\right)~, \cr      }}
and the Casimir becomes%
\eqn\bfx{\eqalign{{\cal H}^2=\p_r\Delta \partial_r -{\left({2Mr_+\p_t}+{a}\p_\phi \right)^2\over (r-r_+)(r_+-r_-)}  +{\left({2Mr_-\p_t}+{a}\p_\phi \right)^2\over (r-r_-)(r_+-r_-)} ~. }}
The near region
wave equation \bfxx\ can be written as
\eqn\mlk{\bar { \cal H}^2\Phi={\cal H}^2 \Phi =\ell(\ell+1)\Phi~.}
We see that the scalar Laplacian has reduced to the $SL(2,R)$ Casimir.
The
$SL(2,R)_L\times SL(2,R)_R$ weights of the field $\Phi $ are
\eqn\wwws{(h_L,h_R)=(\ell,\ell)~.}

From this result one might think that the
solutions of the Kerr wave equation in the near region form $SL(2,R)$ representations.
In fact this is not the case, because the vectors fields \bzg\ - \bsga\ which generate the $SL(2,R)$ symmetries are
not globally defined. They are not periodic under the angular identification
\eqn\be{\phi \sim \phi +2\pi~.}
Thus these symmetries cannot be used to generate new global solutions from old ones.
This can be interpreted as the statement that the $SL(2,R)_L \times SL(2,R)_R$ symmetry is spontaneously broken by the periodic identification of the angular coordinate $\phi$.  Indeed, under the identification \be\
the conformal coordinates are identified as
\eqn\fds{ w^+ \sim e^{4\pi^2 T_R}w^+,~~w^- \sim e^{4\pi^2 T_L}w^-,~~y \sim e^{2\pi^2 (T_L+T_R)}y~.}
This identification is generated by the $SL(2,R)_L\times SL(2,R)_R$ group element
\eqn\rda{e^{-i4\pi^2T_RH_0-i4\pi^2 T_L\bar H_0 }~.}
Hence the $SL(2,R)_L\times SL(2,R)_R$ symmetry is broken down to the  $U(1)_L\times U(1)_R$ subgroup generated by $(\bar H_0, H_0)$.

The situation is somewhat similar to the BTZ black hole in 2+1 gravity, which has a local
 $SL(2,R)_L\times SL(2,R)_R$ isometry which is spontaneously broken by the identification of the angular coordinate $ \phi$.  In that case the symmetry, even though it is broken by the BTZ geometry, is still usefully present in the theory.  In particular, the conformal symmetry still fixes the form of scattering amplitudes and constrains the asymptotic density of states via Cardy's formula.  The  Kerr case is similar, except that the broken  $SL(2,R)_L\times SL(2,R)_R$  acts on the solution space but not on the geometry itself.  Nevertheless, we shall see that powerful constraints from symmetry considerations still apply.


\newsec{CFT Interpretation}
\subsec{Temperature}
The $SL(2,R)_L\times SL(2,R)_R$ symmetries described above generate rigid conformal transformations in the $(w^+, w^-)\sim (\phi,t)$ plane.  Accordingly, let us now assume that the dynamics of the near region is described by a dual 2D CFT, which possesses a ground state that is invariant under the full $SL(2,R)_L\times SL(2,R)_R$ symmetry.  What then is the effect of the identification
\fds? \foot{The analysis here follows that of \MaldacenaBW\ for the BTZ black hole. Although the discussion of \MaldacenaBW\ was in the context of string theory, the discussion did not actually require string theory.}    At fixed $r$,  the relation between conformal $(w^+, w^-)$ and Boyer-Lindquist $(\phi,t)$ coordinates is, up to an $r$-dependent rescaling,
\eqn\csc{\eqalign{w^\pm &= e^{\pm t^\pm}~,}}
with
\eqn\csca{\eqalign{ t^+& ={2\pi T_R \phi}~,\cr t^-&={t \over 2M} -2\pi T_L \phi~.
}}
This is precisely the relation between Minkowski $(w^\pm)$ and Rindler $(t^\pm)$ coordinates. In the $SL(2,R)_L\times SL(2,R)_R$ invariant Minkowski vacuum, observers at fixed position in Rindler coordinates will observe a thermal bath of Unruh radiation.  The periodic identification of $\phi$ requires that we restrict our observations to a fundamental domain of the identification
\eqn\oooou{t^+\sim t^++4\pi^2 T_R~,~~~t^-\sim t^--4\pi^2 T_L~.}
The quantum state describing physics in this accelerating strip of Minkowski space is obtained from the Minkowski vacuum by tracing over the quantum state in the region outside the strip. The well-known result is that we get a thermal density matrix at temperature $(T_L,T_R)$.  Hence the Kerr black hole should be dual to a finite temperature $(T_L,T_R)$ mixed state in the dual CFT.

\subsec{Entropy}
We would now like to microscopically reproduce the Kerr entropy by assuming the Cardy formula for the dual 2D CFT.  This requires a formula for the central charges $c_L $ and $c_R$.
In some cases, such central charges can be derived from an analysis of the asymptotic symmetry group \refs{\ReggeZD, \BrownNW, \BarnichJY, \BarnichBF}. This derivation has been completed for extreme Kerr,
giving  \refs{\GuicaMU, \MatsuoSJ,\CastroJF}
\eqn\tkl{c_R=c_L=12J~.}
So far, as mentioned in the introduction, no one has understood how to extend this calculation beyond linear order away from extremality. In this paper we have adopted an alternate approach which does not lead to a formula for $c_{L,R}$. Therefore we will simply assume that the conformal symmetry found here connects smoothly to that of the extreme limit and that the central charge is therefore still given by \tkl.  The Cardy formula for the microstate degeneracy is
\eqn\ccf{S={\pi^2\over 3}(c_LT_L+c_RT_R)~.}
Using the central charges \tkl\ and temperatures \llj\ we get
\eqn\ccg{S=2\pi Mr_+={{\rm Area} \over 4}~.}
This agrees on the nose with the macroscopic Bekenstein-Hawking area law for the entropy.\foot{A similar derivation was attempted in \CveticJN\ but was missing an overall mutliplicative factor.}$^{,}$\foot{A sufficient condition for validity of  the Cardy formula \ccf\ (in a unitary theory) is that the temperatures $(T_L,T_R)$ are large compared to the central charge.  For suitable choices of parameters this indeed holds.  This includes the near-Schwarzschild case where $M \gg J\ne 0$. Outside this parameter range the applicability of the Cardy formula may still follow, as in stringy examples \masu, in the long string picture.
}


\newsec {Scattering}

If the near region of Kerr is dual to a 2D CFT, then near region contributions to scattering amplitudes or absorption probabilities should be given by 2D CFT two-point functions. We will see in this section that that this is indeed the case. The derivation here is essentially identical to that given many times before staring with
\MaldacenaIH\ and we will accordingly be brief.
The only difference is that in the present context the near region is not geometrically a near-horizon region, but this does not affect the discussion.

\subsec{Absorption Probabilities}
The absorption probability for a massless scalar $\Phi$ at frequencies $\omega M\ll 1$ and arbitrary $m$, $\ell$ was computed long ago \sct\ and re-analyzed in \refs{\MaldacenaIH,\CveticXV,\CveticAP}.   In the near region $\omega r \ll 1$ the solution to the radial wave equation \bbz\  with ingoing boundary conditions at the horizon  is
\eqn\bfa{\eqalign{R(r)=&\left({r-r_+\over r-r_-}\right)^{-i{2Mr_+\over r_+-r_-}(\omega-m\Omega)} (r-r_-)^{-1-\ell}\cr &F\left(1+\ell-i{4M\over r_+-r_-}(M{\omega}-r_+m\Omega),1+\ell-i2M{\omega}; 1-i{4Mr_+\over r_+-r_-}(\omega-m\Omega); {r-r_-\over r-r_+}\right)~,}}
where $F(a,b;c;z)$ is the hypergeometric function and
\eqn\qa{\Omega={a\over 2Mr_+}~,}
is the angular velocity at the horizon.
At the outer boundary of the matching region  $r\gg M$ (but still $r\ll {1 \over\omega}$)  \bfa\ behaves as
\eqn\fb{R(r\gg M)\sim Ar^\ell+ Br^{-1-\ell} \sim A r^\ell~,}
with
\eqn\fba{\eqalign{ A&={\Gamma(1-i{4Mr_+\over r_+-r_-}(\omega-m\Omega))\Gamma(1+2\ell)\over\Gamma(1+\ell-i{2M\omega})\,\Gamma(1+\ell-i{4M^2\over r_+-r_-}\omega+i{4Mr_+\Omega\over r_+-r_-}m)}~,
}}
%
up to an overall constant independent of $\omega$ and $m$. A similar expression for $B$ -- which is not needed here --  can be found in  \BredbergPV.
The absorption cross section is then proportional to
 \eqn\fc{\eqalign{P_{\rm abs}&\sim |A|^{-2}\cr & \sim \sinh\left({4\pi Mr_+\over r_+-r_-}(\omega-m\Omega)\right)\left|\Gamma\left(1+\ell-i{2M\omega}\right)\right|^2\times\cr & \quad\quad \quad\,\left|\Gamma\left(1+\ell-i{4M^2\over r_+-r_-}\omega+i{4Mr_+\Omega\over r_+-r_-}m\right)\right|^2 }}

   To compare with the dual CFT we rewrite $P_{\rm abs}$ in terms of the CFT temperatures $(T_R,T_L)$, the linearization of their conjugate charges and the conformal weights $(\ell,\ell)$. To determine the linearized conjugate charges we begin with  the first law of thermodynamics
\eqn\jjl{T_H\delta S=\delta M-\Omega \delta J,}
where
\eqn\jja{ T_H={1\over 8\pi}{r_+-r_-\over Mr_+}~,}
$S=2\pi Mr_+$, and we identify $\omega=\delta M$ and $m=\delta J$. We then look for the conjugate charges  $\delta E_R$ and $\delta E_L$ such that
\eqn\rrp{ \delta S={\delta E_L \over T_L}+{\delta E_R \over T_R}}
with $T_{L,R}$ given by \llj. The solution is
\eqn\fea{\eqalign{\delta E_L&= {2M^3\over J}\delta M~,\cr \delta E_R &={2M^3\over J}\delta M-\delta J~,}}
hence we identify the left and right moving frequencies as
\eqn\fe{\eqalign{ \omega_L\equiv \delta E_L&= {2M^3\over J}\omega~,\cr  \omega_R \equiv  \delta E_R&={2M^3\over J}\omega-m}~.}
Using these formula as well as \llj\ and \wwws\ one then finds that the gravity result \fc\
can be expressed as
\eqn\fa{P_{\rm abs}\sim T_L^{2h_L-1}T_R^{2h_R-1}\sinh\left({\omega_L\over 2T_L}+{\omega_R \over 2 T_R}\right)\left|\Gamma(h_L+i{\omega_L \over 2\pi T_L})\right|^2\left|\Gamma(h_R+i{\omega_R\over 2\pi T_R})\right|^2~,}
which is precisely the well-known finite-temperature absorption cross section for a 2D CFT.

\subsec{Parameter Ranges}

The nature of the agreement between the CFT and gravity results for $\omega M\ll1$ depends on the values of the parameters under consideration. While \fa\ is the correct gravity answer whenever
$\omega M \ll1$, in some cases the expression \fa\ can be organized into leading and subleading terms. In these cases only the leading term can obviously be trusted and a more detailed analysis is required to see if corrections to the matching procedure effect the result. For this reason, although
the gravity and CFT do agree insofar as they have been tested, the test is not as strong as it may first appear from \fa. For example, for generic values of $M$ which differ from $\sqrt J$ by a multiplicative factor of order unity, $T_L$ and $T_R$ are themselves of order unity.  It follows that ${\omega_L
\over T_L} \ll 1$  while $\omega_R \sim -m$.  The leading term in \fa\ is then
\eqn\faz{P_{\rm abs}\sim - T_L^{2h_L-1}T_R^{2h_R-1}\sinh\left({m \over 2 T_R}\right)\left|\Gamma(h_L)\right|^2\left|\Gamma(h_R-i{m\over 2\pi T_R})\right|^2+{\cal O}(\omega M)~,}
which does not involve $\omega$.
The fact that a (variant of) this  expression  has a CFT interpretation was already noted in \refs{\CveticJN,\MaldacenaIH}.  Another interesting case is when $m=0$ and $T_R$ is of order $M\omega$, from which it follows that ${\omega_R \over T_R}$ is of order one.   In this case the leading order answer does depend on $\omega$ and  we overlap the parameter range considered in \BredbergPV. In this overlap range, the agreement here is equivalent to what was found there.

Hence the results of this paper lend further credence to the idea that there is a general conformal symmetry governing the dynamics of Kerr black holes of which the discussions of \refs{ \MaldacenaIH,\BredbergPV,\GuicaMU, \MatsuoSJ,\CastroJF,\CveticJN} comprise various aspects and special cases. We hope to understand this more completely.

\centerline{\bf Acknowledgements}
This work was in part motivated by comments over the years from Finn Larsen that the appearance of hypergeometric functions in black hole scattering amplitudes indicates a hidden conformal symmetry. We are grateful to Tom Hartman, Finn Larsen, Juan Maldacena and Wei Song for useful conversations. This work was supported in part by DOE grant DE-FG0291ER40654 and the National Science and Engineering Research Council of Canada.
\listrefs

\end